# Biological Impact on Military Intelligence: Application or Metaphor?

Lester Ingber

Lester Ingber Research, Ashland OR, USA

ingber@alumni.caltech.edu   http://www.ingber.com

## Abstract

Ideas by Statistical Mechanics (ISM) is a generic program to model evolution and propagation of ideas/patterns throughout populations subjected to endogenous and exogenous interactions. The program is based on the author's work in Statistical Mechanics of Neocortical Interactions (SMNI). This product can be used for decision support for projects ranging from diplomatic, information, military, and economic (DIME) factors of propagation/evolution of ideas, to commercial sales, trading indicators across sectors of financial markets, advertising and political campaigns, etc. It seems appropriate to base an approach for propagation of ideas on the only system so far demonstrated to develop and nurture ideas, i.e., the neocortical brain. The issue here is whether such biological intelligence is a valid application to military intelligence, or is it simply a metaphor?

---

# 1. Introduction

There is a fairly large literature, a small segment referenced below, that considers application of biological intelligence (BI) to military intelligence (MI). The query naturally arises as to whether this is a *bona fide* application BI to MI, or is this a metaphor?

Of course, if BI application is a metaphor, this still could be pedagogically useful is some tactical or strategic or political decisions impacting $C^3I$ processes, but if this is only a metaphor, then beware of improper applications.

This paper summarizes some recent work in understanding human brain processing of information, involving multiple scales of interaction — bottom-up as well as top-down — from quantum and molecular scales up to scales involving synchronous firing of many neurons large enough to be measured as scalp Electroencephalographic (EEG), i.e., interactions among scales ranging from $10^{-8}$m to $10^{-2}$m, a range of six orders of magnitude.

This presents quite a different context of the brain than a simple artificial neural network that simply considers the development of patterns of information as arising form the bottom-up substrate of interaction (quasi-)neurons, and which does not include the direct top-down influences of the largest scales of activity directly on multiple smaller scales.

The body of this paper deals with neocortex. The issue here is whether there is any *bona fide* application, not just metaphor, to MI. We are concerned with human interactions that define combat and war and politics, so it makes some sense to examine this issue (Ingber & Sworder, 1991). For example, can "leadership" be effectly imparted directly down to individual troops, without directly involving chain of command, and if so is this an example of such an application of BI? If we think so, can anything be learned from real neocortex? This paper does not answer these questions; it just raises the issue in the context of including real BI.

## 1.1. DIME Factors

A briefing (Allen, 2004) demonstrates the breadth and depth complexity required to address real diplomatic, information, military, economic (DIME) factors for the propagation/evolution of ideas through defined populations. An open mind would conclude that it is possible that multiple approaches may be required for multiple decision makers in multiple scenarios. However, it is in the interests of multiple decision–makers to as much as possible rely on the same generic model for actual computations. Many users would have to trust that the coded model is faithful to process their inputs.

The concept of "memes" is an example of an approach to deal with DIME factors (Situngkir, 2004). The meme approach, using a reductionist philosophy of evolution among genes, is reasonably contrasted to approaches emphasizing the need to include relatively global influences of evolution (Thurtle, 2006).

There are multiple other alternative works being conducted world–wide that must be at least kept in mind while developing and testing models of evolution/propagation of ideas in defined populations: A study on a simple algebraic model of opinion formation concluded that the only final opinions are extremal ones (Aletti *et al*, 2006). A study of the influence on chaos on opinion formation, using a simple algebraic model, concluded that contrarian opinion could persist and be crucial in close elections, albeit the authors were careful to note that most real populations probably do not support chaos (Borghesi & Galam, 2006). A limited review of work in social networks illustrates that there are about as many phenomena to be explored as there are disciplines ready to apply their network models (Sen, 2006).

## 1.2. Multiple Scales of Neocortical Interactions

There is a growing awareness of the importance of multiple scales in many physical and biological systems, including neuroscience (Anastassiou *et al*, 2011; Nunez *et al*, 2013). As yet, there do not seem to be any explicit top-down mechanisms that directly drive bottom-up processes that describe memory, attention, etc. Of course, there are many top-down type studies demonstrating that neuromodulator (Silberstein, 1995) and neuronal firing states, e.g., as defined by EEG frequencies, can modify the milieu of individual synaptic and neuronal activity, which is still consistent with ultimate bottom-up paradigms. However, there is a logical difference between top-down milieu as conditioned by some prior external or internal conditions, and some direct top-down processes that direct cause bottom-up interactions specific to short-term memory (STM).

A recent study crosses molecular ($Ca^{2+}$ ions), microscopic (synaptic and neuronal), mesoscopic (minicolumns and macrocolumns), and macroscopic (regional scalp EEG) scales (Ingber, Pappalepore & Stesiak, 2014; Ingber, 2014b). Calculations support the interaction between synchronous columnar firings large enough to be measured by scalp EEG and molecular scales contributing to synaptic activity: On one hand, the influence of macroscopic scales on molecular scales is calculated via the evolution of $Ca^{2+}$ quantum wave functions. On the other hand, the influence of $Ca^{2+}$ waves is described in the context of a statistical mechanics model that already has been verified as calculating experimental observables, aggregating and scaling up from synaptic activity, to columnar neuronal firings, to regional synchronous activity fit to EEG while preserving an audit trail back to underlying synaptic interactions.

In the above context of multiple scales of neocortical interactions, it seems reasonable to propose that an Artificial Intelligence (AI)/robotic system that wishes to take advantage of modelling of neocortex take such multiple scales of interaction into account in basic design.

## 1.3. Statistical Mechanics of Neocortical Interactions (SMNI)

A class of AI algorithms that has not yet been developed in this context takes advantage of information known about real neocortex. It seems appropriate to base an approach for propagation of ideas on the only system so far demonstrated to develop and nurture ideas, i.e., the neocortical brain. A statistical mechanical model of neocortical interactions, developed by the author and tested successfully in describing short–term memory (STM) and EEG indicators, is the proposed bottom–up model. evolution and propagation of ideas/patterns throughout populations subjected to endogenous and exogenous interactions (Ingber, 2006).

Ideas by Statistical Mechanics (ISM) develops subsets of macrocolumnar activity of multivariate stochastic descriptions of defined populations, with macrocolumns defined by their local parameters within specific regions and with parameterized endogenous inter–regional and exogenous external connectivities. Parameters of subsets of macrocolumns will be fit to patterns representing ideas. Parameters of external and inter–regional interactions will be determined that promote or inhibit the spread of these ideas. Fitting such nonlinear systems requires the use of sampling techniques.

The author's approach uses guidance from his statistical mechanics of neocortical interactions (SMNI), developed in a series of about 30 published papers from 1981–2014 (Ingber, 1983; Ingber, 1985; Ingber, 1992; Ingber, 1994; Ingber, 1995; Ingber, 1997; Ingber, 2011; Ingber, 2012; Nunez *et al*, 2013; Ingber, Pappalepore & Stesiak, 2014). These papers also address long–standing issues of information measured by EEG as arising from bottom–up local interactions of clusters of thousands to tens of thousands of neurons interacting via short–ranged fibers), or top–down influences of global interactions (mediated by long–ranged myelinated fibers). SMNI does this by including both local and global interactions as being necessary to develop neocortical circuitry.

## 1.4. Executive Summary: "Mind Over Matter"

"Mind Over Matter" is a stretch, but not inaccurate, context. The logic is:

(1) A lot of what we consider "mind" is conscious attention to short-term memories, which can develop by (a) external stimuli directly, (b) internal long-term storage, (c) new ideas=memories developed in abstract regions of the brain, etc.

(2) It is now accepted by some to many neuroscientists (e.g., see the 2012 National Institute of Health (NIH) news article http://goo.gl/cxNOa0), and confirmed by some experiments, that at least some such memories in (1) are actively processed by highly synchronized patterns of neuronal firings, with enough synchrony to be able to be easily measured by scalp EEG recordings during activity of processing such patterns, e.g., P300 waves, etc. These minicolumnar currents giving rise to measurable EEG also give rise to magnetic vector potentials **A**. The **A** fields have a logarithmic range insensitivity and are additive over larger distances than electric **E** or magnetic **B** fields.

(3) A recent Journal of Theoretical Biology (JTB) paper (Ingber, Pappalepore & Stesiak, 2014) calculates the influence of such synchronous EEG at molecular scales of $Ca^{2+}$ ionic waves, a process which is present in the body and brain, but particularly as astrocyte influences at synaptic gaps, thereby affecting background synaptic activity, which in turn can be synchronized by other processes to give rise to the large-scale activity discussed in (1). The $Ca^{2+}$ wave have a duration of momentum **p** which is observed to be rather large.

(4) The recent paper connects the influence of (1) over (3) directly via a specific interaction, $\mathbf{p} + q\mathbf{A}$, where $q$ for a $Ca^{2+}$ ion $= 2e$, where $e$ is the magnitude of the charge of an electron. The $\mathbf{p} + q\mathbf{A}$ interaction is well established in both classical and quantum physics.

The direct $\mathbf{p} + q\mathbf{A}$ influence of (1) over (3) can reasonably be discussed as a "mind over matter" process. E.g., just thinking about thinking can give rise to this effect.

(5) The JTB paper reports the use of MPI to parallelize code that fits EEG data to a model that includes the dynamic influence of a $\mathbf{p} + q\mathbf{A}$ interaction on background synaptic activity, embedded in my Statistical Mechanics of Neocortical Interactions (SMNI) coded model, to assess the viability EEG activity influencing this interaction. This now provides a testbed for future models of such interactions.

## 2. SMNI Applied to Artificial Intelligence

The application of SMNI to AI has been discussed in previous papers (Ingber, 2007; Ingber, 2008; Ingber, 2014a). Neocortex has evolved to use minicolumns of neurons interacting via short−ranged interactions in macrocolumns, and interacting via long−ranged interactions across regions of macrocolumns. This common architecture processes patterns of information within and among different regions of sensory, motor, associative cortex, etc. Therefore, the premise of this approach is that this is a good model to describe and analyze evolution/propagation of Ideas among defined populations.

Relevant to this study is that a spatial−temporal lattice−field short−time conditional multiplicative−noise (nonlinear in drifts and diffusions) multivariate Gaussian−Markovian probability distribution is developed faithful to neocortical function/physiology. Such probability distributions are a basic input into the approach used here. The SMNI model was the first physical application of a nonlinear multivariate calculus developed by other mathematical physicists in the late 1970's to define a statistical mechanics of multivariate nonlinear nonequilibrium systems (Graham, 1977; Langouche *et al*, 1982).

## 2.1. SMNI Tests on STM and EEG

SMNI builds from synaptic interactions to minicolumnar, macrocolumnar, and regional interactions in neocortex. Since 1981, a series of SMNI papers has been developed model columns and regions of neocortex, spanning mm to cm of tissue. Most of these papers have dealt explicitly with calculating properties of STM and scalp EEG in order to test the basic formulation of this approach (Ingber, 1983; Ingber, 1985; Ingber & Nunez, 1995).

The SMNI modeling of local mesocolumnar interactions (convergence and divergence between minicolumnar and macrocolumnar interactions) was tested on STM phenomena. The SMNI modeling of macrocolumnar interactions across regions was tested on EEG phenomena.

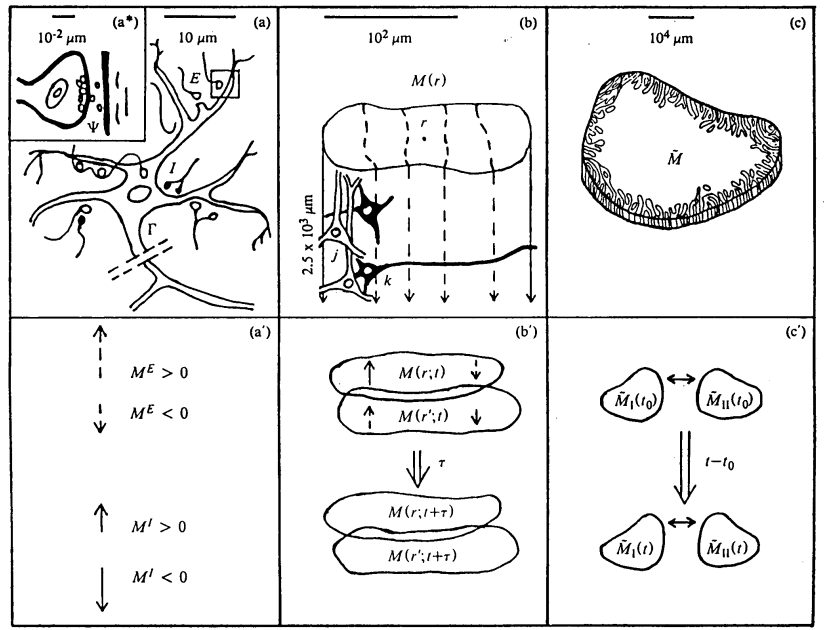

Figure 1. Illustrated are three biophysical scales of neocortical interactions: (a)–(a*)–(a') microscopic neurons; (b)–(b') mesocolumnar domains; (c)–(c') macroscopic regions (Ingber, 1983). SMNI has developed appropriate conditional probability distributions at each level, aggregating up from the smallest levels of interactions. In (a*) synaptic inter–neuronal interactions, averaged over by mesocolumns, are phenomenologically described by the mean and variance of a distribution $\Psi$. Similarly, in (a) intraneuronal transmissions are phenomenologically described by the mean and variance of $\Gamma$. Mesocolumnar averaged excitatory ($E$) and inhibitory ($I$) neuronal firings $M$ are represented in (a'). In (b) the vertical organization of minicolumns is sketched together with their horizontal stratification, yielding a physiological entity, the mesocolumn. In (b') the overlap of interacting mesocolumns at locations $r$ and $r'$ from times $t$ and $t + \tau$ is sketched. In (c) macroscopic regions of neocortex are depicted as arising from many mesocolumnar domains. (c') sketches how regions may be coupled by long–ranged interactions.

## 2.2. SMNI Description of STM

SMNI studies have detailed that maximal numbers of attractors lie within the physical firing space of both excitatory and inhibitory minicolumnar firings, consistent with experimentally observed capacities of auditory and visual STM, when a "centering" mechanism is enforced by shifting background noise in synaptic interactions, consistent with experimental observations under conditions of selective attention (Ingber, 1985; Ingber, 1994).

These calculations were further supported by high−resolution evolution of the short−time conditional−probability propagator using PATHINT (Ingber & Nunez, 1995). SMNI correctly calculated the stability and duration of STM, the primacy versus recency rule, random access to memories within tenths of a second as observed, and the observed $7 \pm 2$ capacity rule of auditory memory and the observed $4 \pm 2$ capacity rule of visual memory.

SMNI also calculates how STM patterns (e.g., from a given region or even aggregated from multiple regions) may be encoded by dynamic modification of synaptic parameters (within experimentally observed ranges) into long−term memory patterns (LTM) (Ingber, 1983).

PATHINT also has been used to calculate models of National Training Center (NTC) data (Ingber, Fujio & Wehner, 1991; Ingber, 1993; Ingber, 2001).

## 2.3. SMNI Description of EEG

Using the power of this formal structure, sets of EEG and evoked potential data from a separate NIH study, collected to investigate genetic predispositions to alcoholism, were fitted to an SMNI model on a lattice of regional electrodes to extract brain "signatures" of STM (Ingber, 1997). Each electrode site was represented by an SMNI distribution of independent stochastic macrocolumnar−scaled firing variables, interconnected by long−ranged circuitry with delays appropriate to long-fiber communication in neocortex. The global optimization algorithm ASA was used to perform maximum likelihood fits of Lagrangians defined by path integrals of multivariate conditional probabilities. Canonical momenta indicators (CMI) were thereby derived for individual's EEG data. The CMI give better signal recognition than the raw data, and were used to advantage as correlates of behavioral states. In−sample data was used for training (Ingber, 1997), and out−of−sample data was used for testing these fits.

The architecture of ISM is modeled using scales similar to those used for local STM and global EEG connectivity.

CMI have also been used in the context of nonlinear combat using training data (Bowman & Ingber, 1997).

## 2.4. Generic Mesoscopic Neural Networks

SMNI was applied to a parallelized generic mesoscopic neural networks (MNN) (Ingber, 1992), adding computational power to a similar paradigm proposed for target recognition.

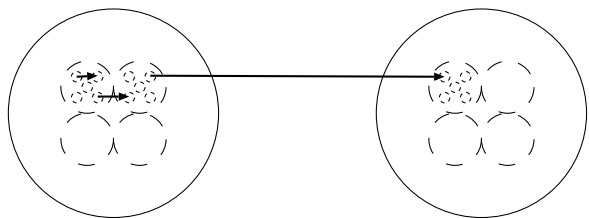

Figure 2. Scales of interactions among minicolumns are represented, within macrocolumns, across macrocolumns, and across regions of macrocolumns.

"Learning" takes place by presenting the MNN with data, and parametrizing the data in terms of the firings, or multivariate firings. The "weights," or coefficients of functions of firings appearing in the drifts and diffusions, are fit to incoming data, considering the joint "effective" Lagrangian (including the logarithm of the prefactor in the probability distribution) as a dynamic cost function. This program of fitting coefficients in Lagrangian uses methods of ASA.

"Prediction" takes advantage of a mathematically equivalent representation of the Lagrangian path−integral algorithm, i.e., a set of coupled Langevin rate−equations. A coarse deterministic estimate to "predict" the evolution can be applied using the most probable path, but PATHINT has been used. PATHINT, even when parallelized, typically can be too slow for "predicting" evolution of these systems. However, PATHTREE is much faster (Ingber, Chen *et al*, 2001).

**Architecture for Selected ISM Model**.

The primary objective is to deliver a computer model that contains the following features: (1) A multivariable space will be defined to accommodate populations. (2) A cost function over the population variables in (1) will be defined to explicitly define a pattern that can be identified as an Idea. A very important issue is for this project is to develop cost functions, not only how to fit or process them. (3) Subsets of the population will be used to fit parameters — e.g, coefficients of variables, connectivities to patterns, etc. — to an Idea, using the cost function in (2). (4) Connectivity of the population in (3) will be made to the rest of the population. Investigations will be made to determine what endogenous connectivity is required to stop or promote the propagation of the Idea into other regions of the population. (5) External forces, e.g., acting only on specific regions of the population, will be introduced, to determine how these exogenous forces may stop or promote the propagation of an Idea.

**Multiple Scales of SMNI Interactions**.

A model has been developed to calculate and experimentally test the coupling of molecular scales of $Ca^{2+}$ wave dynamics with **A** fields developed at macroscopic regional scales measured by coherent neuronal firing activity measured by scalp EEG (Ingber, Pappalepore & Stesiak, 2014). The author is PI of a 2013-2014 computer grant that made this work possible, under the National Science Foundation Extreme Science and Engineering Discovery Environment (XSEDE.org). See http://www.ingber.com/lir_computational_physics_group.html for information for volunteers.

For several decades biological and biophysical research into neocortical information processing has explained neocortical interactions as specific bottom-up molecular and smaller-scale processes (Rabinovich *et al*, 2006). It is clear that most molecular approaches consider it inevitable that their approaches at molecular and possibly even quantum scales will yet prove to be causal explanations of relatively macroscopic phenomena.

This recent study crosses molecular, microscopic (synaptic and neuronal), mesoscopic (minicolumns and macrocolumns), and macroscopic regional scales. Over the past three decades, with regard to STM and LTM phenomena, which themselves are likely components of other phenomena like attention and consciousness, the SMNI approach has yielded specific details of STM not present in molecular approaches (Ingber, 2012). The SMNI calculations detail information processing capable of neocortex using patterns of columnar firings, e.g., as observed in scalp EEG (Salazar *et al*, 2012), which give rise to a SMNI vector potential **A** that influences the molecular $Ca^{2+}$ momentum **p**, and thereby synaptic interactions. Explicit Lagrangians have been given, serving as cost/objective functions that can be fit to EEG data using ASA, as similarly performed in previous SMNI papers (Ingber, 1997; Ingber, 1998).

Considerations in both classical and quantum physics predict a predominance of $Ca^{2+}$ waves in directions closely aligned to the direction perpendicular to neocortical laminae (**A** is in the same direction as the current flow, typically across laminae, albeit they are convoluted), especially during strong collective EEG (e.g., strong enough to be measured on the scalp, such as during selective attention tasks). Since the spatial scales of $Ca^{2+}$ wave and macro-EEG are quite disparate, an experimenter would have to be able to correlate both scales in time scales on the order of tens of milliseconds.

The basic premise of this study is robust against much theoretical modeling, as experimental data is used wherever possible for both $Ca^{2+}$ ions and for large-scale electromagnetic activity. The theoretical construct of the canonical momentum $\Pi = \mathbf{p} + q\mathbf{A}$ is firmly entrenched in classical and quantum mechanics. Calculations demonstrate that macroscopic EEG $\mathbf{A}$ can be quite influential on the momentum $\mathbf{p}$ of $Ca^{2+}$ ions, at scales of both classical and quantum physics.

A single $Ca^{2+}$ ion can have a momentum appreciably altered in the presence of macrocolumnar EEG firings, and this effect is magnified when many ions in a wave are similarly affected. Therefore, large-scale top-down neocortical processing giving rise to measurable scalp EEG can directly influence molecular-scale bottom-up processes. This suggests that, instead of the common assumption that $Ca^{2+}$ waves contribute to neuronal activity, they may in fact at times be caused by the influence of $\mathbf{A}$ of larger-scale EEG. The SMNI model supports a mechanism wherein the $\mathbf{p} + q\mathbf{A}$ interaction at tripartite synapses, via a dynamic centering mechanism (DCM) to control background synaptic activity, acts to maintain STM during states of selective attention. Such a top-down effect awaits forensic in vivo experimental verification, requiring appreciating the necessity and due diligence of including true multiple-scale interactions across orders of magnitude in the complex

**Application of SMNI Model**.

The approach is to develop subsets of Ideas/macrocolumnar activity of multivariate stochastic descriptions of defined populations (of a reasonable but small population samples, e.g., of 100–1000), with macrocolumns defined by their local parameters within specific regions (larger samples of populations) and with parameterized long–ranged inter–regional and external connectivities. Parameters of a given subset of macrocolumns will be fit using ASA to patterns representing Ideas, akin to acquiring hard–wired long–term memory (LTM) patterns. Parameters of external and inter–regional interactions will be determined that promote or inhibit the spread of these Ideas, by determining the degree of fits and overlaps of probability distributions relative to the seeded macrocolumns.

That is, the same Ideas/patterns may be represented in other than the seeded macrocolumns by local confluence of macrocolumnar and long–ranged firings, akin to STM, or by different hard–wired parameter LTM sets that can support the same local firings in other regions (possible in nonlinear systems). SMNI also calculates how STM can be dynamically encoded into LTM (Ingber, 1983).

Small populations in regions will be sampled to determine if the propagated Idea(s) exists in its pattern space where it did exist prior to its interactions with the seeded population. SMNI derives nonlinear functions as arguments of probability distributions, leading to multiple STM, e.g., $7 \pm 2$ for auditory memory capacity. Some investigation will be made into nonlinear functional forms other than those derived for SMNI, e.g., to have capacities of tens or hundreds of patterns for ISM.

## 3. Conclusion

It seems appropriate to base an approach for propagation of generic ideas on the only system so far demonstrated to develop and nurture ideas, i.e., the neocortical brain. A statistical mechanical model of neocortical interactions, developed by the author and tested successfully in describing short–term memory and EEG indicators, Ideas by Statistical Mechanics (ISM) (Ingber, 2006; Ingber, 2007) is the proposed model. ISM develops subsets of macrocolumnar activity of multivariate stochastic descriptions of defined populations, with macrocolumns defined by their local parameters within specific regions and with parameterized endogenous inter–regional and exogenous external connectivities.

The issue raised here is whether Biological Intelligence (BI), as modeled by Statistical Mechanics of Neocortical Interactions (SMNI) of the real brain, as mapped into Ideas by Statistical Mechanics (ISM), is a reasonable application to Military Intelligence (MI).